\documentclass[11pt,a4paper]{article}
\usepackage{amsmath,amssymb,bm}
 \usepackage{epsfig,graphicx}
  \usepackage{cite}
   \usepackage{pifont}
\newcommand{\nn}{\nonumber}
\newcommand{\Ds}{\displaystyle}

\begin{document}
\title{\bf End-point behavior of the pion distribution amplitude
       \footnote{
                 Invited talk given by the second author at
                 "QUARKS-2010, 16th International Seminar on High Energy Physics",
                 Kolomna, Russia, 6-12 June, 2010.
                 }
      }
\author{
    S. V. Mikhailov$^a$\footnote{{\bf e-mail}: mikhs@theor.jinr.ru},
    A. V. Pimikov$^a$\footnote{{\bf e-mail}: pimikov@theor.jinr.ru},
    N. G. Stefanis$^b$\footnote{{\bf e-mail}: stefanis@tp2.ruhr-uni-bochum.de}
\\
$^a$ \small{\em Bogoliubov Laboratory of Theoretical Physics,
                JINR, 141980 Dubna, Moscow region, Russia} \\
$^b$ \small{\em Institut f\"{u}r Theoretische Physik II,
                Ruhr-Universit\"{a}t Bochum, D-44780 Bochum, Germany} \\
}
\date{}
\maketitle

\begin{abstract}
We discuss the end-point behavior of the pion distribution
amplitude (DA) and calculate its slope using QCD sum rules with
nonlocal condensates.
This is done in terms of the standard derivative and also with the
help of an ``integral derivative'', recently obtained by us.
Our approach favors a value of the slope of the order (or less) of
the asymptotic DA and is in clear disagreement with flat-type pion DAs.
 \end{abstract}

\section{Introduction}
\label{sec:intro}

The main technique to analyze hard exclusive processes within QCD,
is provided by the factorization of the underlying dynamics into a
hard and a soft part.
The hard part forms the  partonic amplitude of the subprocess at a
large value of the momentum transfer and is amenable to QCD
perturbation theory.
The soft part depends on the distribution amplitude of the hadron(s)
and contains the dynamics at typical hadronic scales; it has,
therefore, to be determined by nonperturbative methods or be extracted
from experimental data.
The collinear factorization applied to the transition form factor (FF)
of two far off-shell photons to a pion leads to the convolution of
these two parts, which, at the leading order of twist two, reads
($\bar{x}\equiv 1-x$)
\begin{equation}
  F^{\gamma^{*}\gamma^{*}\pi}(Q^{2},q^{2})
=
  \frac{\sqrt{2}}{3}f_{\pi}
  \int_{0}^{1}
  \frac{1}{Q^{2}\bar{x} + q^{2}x}~\varphi_{\pi}(x)\,dx\,
  + {\cal O}(1/Q^{4})
\label{eq:leading-FF-term}
\end{equation}
modulo twist-four terms, ignored here.
The main ingredient of the above equation is the pion DA
$\varphi_{\pi}(x)$
which encodes all unknown binding effects of the pion state.
At the considered level of twist two, it is defined by the
following universal matrix element \cite{Rad77}
\begin{equation}
  \langle 0 |\bar{d}(z) \gamma^{\mu}\gamma_{5} [z,0] u(0)
            | \pi (P)
  \rangle |_{z^{2}=0}
=
   i f_{\pi}P^{\mu} \int_{0}^{1} {\rm e}^{ix(z\cdot P)}
   \varphi_{\pi}^{\rm (t=2)}(x,\mu_{0}^{2})\,dx \ ,
\label{eq:pi-DA}
\end{equation}
where $x$ is the longitudinal momentum fraction carried by the
valence quark ($\bar{x}$ for the antiquark) in the pion and the
path-ordered exponential, i.e., the light-like gauge link,
\begin{equation}
  [z,0]
=
  {\cal P} \exp \left[ -ig \int_{0}^{z}\!t^{a}A_{\mu}^{a}(y)\,dy^{\mu}
                \right] \, ,
\label{eq:con}
\end{equation}
ensures gauge invariance.
It is useful to expand the pion DA in terms of the Gegenbauer harmonics
$6x\bar{x}~C_{n}^{3/2}(2x-1)$
which provide the eigenfunctions of the one-loop
Efremov-Radyushkin-Brodsky-Lepage (ERBL for short) evolution equation
\cite{ER80tmf,LB80}.
One finds for $\varphi^{\rm (t=2)}$ at the typical hadronic scale
$\mu_0^2$:
\begin{equation}
   \varphi^{\rm (t=2)}(x; \mu_0^2)
=
   \varphi^{\rm Asy}(x)
                      \left[1 + a_2(\mu_0^2)\ C^{3/2}_2(2x-1)
   + a_4(\mu_0^2)\ C^{3/2}_4(2x-1)
       + \ldots  \right]\ ,
\label{eq:pion-DA}
\end{equation}
in which the asymptotic (abbreviated by Asy) pion DA appears:
$\varphi^{\rm Asy}(x)=6x\bar{x}$.
By virtue of the leptonic decay
$\pi\to\mu^{+}\nu_{\mu}$, one obtains the normalization
$\int_{0}^{1}\varphi_{\pi}^{\rm (t=2)}(x,\mu_{0}^{2})\,dx=1$,
which fixes $a_0=1$.

While a process involving two photons with large virtualities is
theoretically preferable, because one can safely apply QCD perturbation
theory, experimentally, the asymmetric kinematic with one of the
photons being quasi real is more accessible.
Indeed, such measurements have been carried out by several
collaborations, namely, the CELLO~\cite{CELLO91},
the CLEO\cite{CLEO98}, and, most recently, the BaBar Collaboration
\cite{BaBar09}.
Taking the limit $q^2\to 0$ in convolution (\ref{eq:leading-FF-term}),
one finds that the FF for the $\gamma^*\gamma\to \pi^0$ transition is
actually given by the inverse moment of the pion DA
\begin{equation}
  \langle x^{-1} \rangle_{\pi}
=
  \int_{0}^{1}\frac{1}{x} \ \varphi_{\pi}(x)\, dx \ .
\label{eq:inv-mom}
\end{equation}
Therefore, this quantity is one of the key elements of the pion-photon
transition FF.
Because this form factor has such a simple structure within QCD, it has
attracted over the years the attention of many theorists
(see, e.g., \cite{RR96,KR96,SSK00,BMS01,MS09} and references cited 
therein).

\begin{figure}[h!]
\centerline{\includegraphics[width=0.5\textwidth]{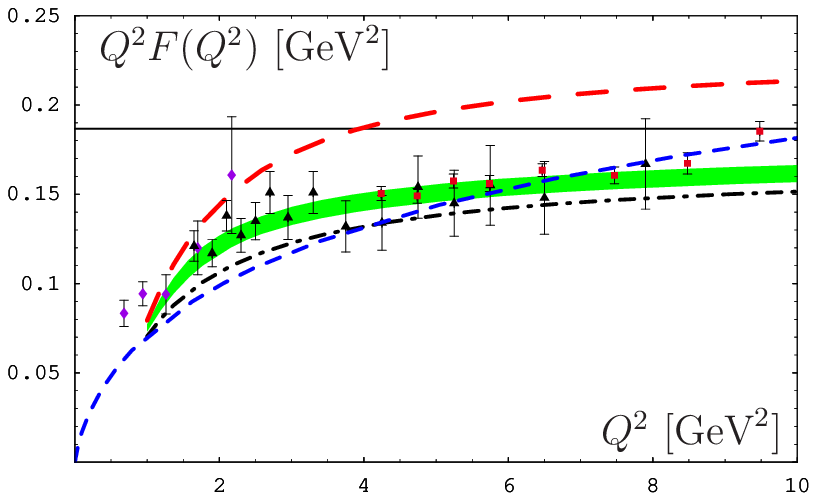}   
           ~\includegraphics[width=0.5\textwidth]{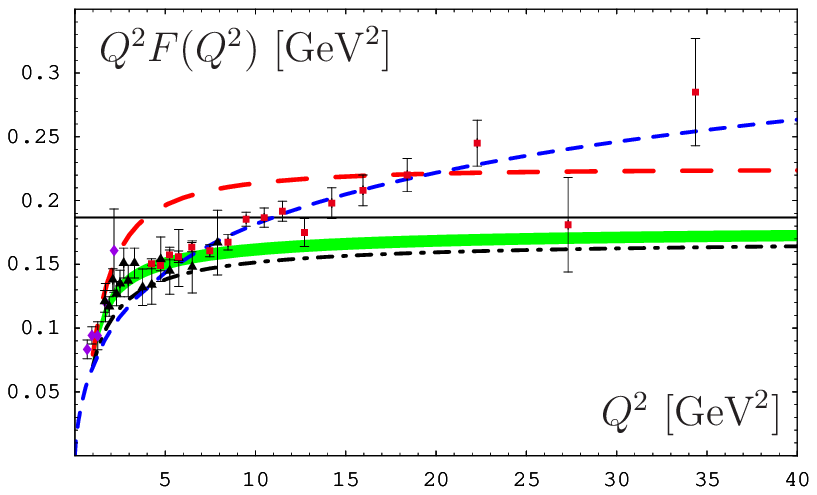}    
           }
\caption{\footnotesize
   Predictions for the photon-to-pion transition FF calculated in
   next-to-leading order (NLO) light-cone sum rules using a Breit-Wigner
   model for the meson resonances.
   The pion is parameterized in terms of the following DAs:
   Asy                                      --- dashed-doted line,
   BMS ``bunch''~\protect\cite{BMS01}       --- shaded (green) strip,
   CZ model~\protect\cite{CZ84}             --- upper long-dashed (red) line,
   flat-top model~\protect\cite{Rad09}      --- short-dashed (blue) line.
   Experimental data are shown on both panels using the following notations:
   BaBar data~\cite{BaBar09}                --- (red) boxes with error bars,
   CELLO data~\cite{CELLO91}                --- (black) diamonds with error bars,
   CLEO data~\cite{CLEO98}                  --- (violet) triangles with error bars.
   The horizontal solid line denotes the asymptotic QCD prediction $\sqrt{2}f_\pi$.
   The left panel provides a zoom-in view into the low and intermediate range of
   $Q^2$, whereas the right panel shows the results for the whole interval of
   momenta probed by the BaBar experiment.
        }
   \label{fig:exp.data.BMS.CZ.Rad.Asy}
\end{figure}

However, the most recent measurement of this observable by the BaBar
Collaboration \cite{BaBar09} has provided controversial results,
because, unexpectedly, the high-energy data points above 10~GeV$^2$
grow with $Q^2$ --- see Fig.\ \ref{fig:exp.data.BMS.CZ.Rad.Asy}.
%
%
At moderate values of the momentum transfer, up to 9~GeV$^2$
(see the left panel of Fig.\ \ref{fig:exp.data.BMS.CZ.Rad.Asy}),
the new high-precision BaBar data agree well with the previous CLEO
data \cite{CLEO98}.
From the second column of Table \ref{tab:all-xi-crit} and the left
panel of Fig.\ \ref{fig:exp.data.BMS.CZ.Rad.Asy}, we may conclude that
all data up to 9~GeV$^2$ can be best described by pion DAs that have
their end-points strongly suppressed
\cite{MS09,MS09Trento_MPL}.
A characteristic example of such a DA is provided by the
Bakulev-Mikhailov-Stefanis (BMS) model \cite{BMS01}, which has been
derived from QCD sum rules (SR)s with nonlocal condensates (NLC)s,
originally developed in \cite{MR89}.
In contrast, the high-$Q^2$ BaBar data
(see the right panel of Fig.\ \ref{fig:exp.data.BMS.CZ.Rad.Asy})
show, as already mentioned, an unexpected growth with $Q^2$ which
cannot be understood on the basis of the collinear factorization and
calls (see the fifth column of Table \ref{tab:all-xi-crit})) for pion
DAs that have instead their end-points strongly enhanced 
\cite{Rad09,Pol09}.
This intriguing behavior has triggered the use of flat-type pion DAs
and provided the main motivation for our analysis in \cite{MPS10},
on which we report here.
The result obtained by Radyushkin \cite{Rad09} with the
flat-top model ($\varphi_\pi(x)=1$) is shown in
Fig.\ \ref{fig:exp.data.BMS.CZ.Rad.Asy} as a dashed (blue) line
in comparison with the BMS bunch (green strip)~\cite{BMS01} and the CZ
pion DA model (long-dashed line in red color)~\cite{CZ84}.
As we see from this figure, the high-$Q^2$ BaBar data are rather
well-described in Radyuskin's approach \cite{Rad09}.
A recent independent analysis \cite{RRBGT10} comes to the conclusion 
that a flat pion DA, when used in a fully consistent way, yields to
predictions for the pion's electromagnetic and transition form factors
which are in striking disagreement with experiment.

\newcommand{\pha}{\strut$\vphantom{\vbox to 5mm{}}\vphantom{_{\vbox to 4mm{}}}$}         
\begin{table}[thb]\vspace*{-3mm}
\caption{\footnotesize
Deviations of theoretical predictions for the quantity
$Q^2F^{\gamma^{*}\gamma\pi}(Q^2)$
in terms of
$\bar{\chi}^2\equiv\chi^2/{\rm ndf}$
(ndf~$=$~number of degrees of freedom)
for several pion model DAs:
Asy, BMS model, and CZ model (more details in \cite{MS09}).
Predictions based on the flat-top model DA (\ref{eq:flat-DA}),
discussed in \cite{Rad09}, are also included.
The second column shows the results for the combined sets of the CLEO
\cite{CLEO98} and the CELLO \cite{CELLO91} data above $1$~GeV$^2$.
The third column compares model predictions with all data in
the interval $[1,9]$~GeV$^2$, while the fourth column takes into
account only the BaBar data above $9$~GeV$^2$.
The last two columns show the values of the inverse moment and the
standard derivative of selected pion DAs at the origin.
\label{tab:all-xi-crit}
}\vspace*{+3mm}
\centerline{
\begin{tabular}{|c||c|c|c||c|c|} \hline
\pha DA Model         & CLEO\&CELLO$$
                                & All data$_{Q^2<9~\text{GeV}^2}$
                                         &  BaBar$_{Q^2>9~\text{GeV}^2}$                                  
                                                      & $\langle x^{-1} \rangle_{\pi}$
                                                                & $\varphi_\pi'(0)$  \\ \hline
\pha Asy              & $2.7$   & $6.35$    & $18.9$  & 3       & 6                   \\ \hline              
\pha BMS~\cite{BMS01} & $0.56$  & $0.86$    & $11.7$  & 3.15    & 1.12                 \\ \hline             
\pha CZ~\cite{CZ84}   & $5.9$   & $33.9$    & $7.9$   & 4.5     & 26                    \\ \hline            
\pha DA~\cite{Rad09}  & $4.15$  & $4.15$    & $1.0$   & ---     & ---                    \\ \hline           
\end{tabular}
}
\end{table}

In our recent investigation \cite{MPS10}, we revisited the QCD SR
approach of \cite{BMS01} focusing our attention on the behavior
of the pion DA in the end-point region $x \sim 0$ with the aim to
understand the fine structure of the pion DA in this region
vs the ansatz for the quark-virtuality distribution in the
nonperturbative QCD vacuum.
As we shall explain in Sec.\ \ref{subsec:int-SR}, QCD SRs were
mainly developed with the purpose to study the \emph{integral}
characteristics of the pion DA.
To overcome this restriction, we constructed in \cite{MPS10}
an operator for \emph{integral derivatives} of the pion DA.
The results obtained this way supplement those found with
SRs which employ the \emph{standard derivative} of the pion DA.

Our presentation will concentrate on the following issues, organized
in sections.
In Sec.\ \ref{sec:pi-slope}, we discuss the QCD SR approach with NLCs
and focus on the peculiarities of the pion DA in the end-point region
$x \sim 0$.
In Sec.\ \ref{subsec:int-SR}  we define and study the integral
derivative of the pion DA, the purpose being to overcome the
restrictions inherent in QCD SRs which were created in order to probe
the \emph{integral} characteristics of the pion DA.
In Sec.\ \ref{subsec:dif-SR} we also study the SRs for the standard
derivative of the pion DA at the origin and complete our presentation
by drawing our conclusions in Sec.\ \ref{sec:concl}.

\section{Slope of the pion DA and the nonperturbative QCD vacuum}
\label{sec:pi-slope}

\begin{figure}[h!]
\centerline{\includegraphics[width=0.45\textwidth]{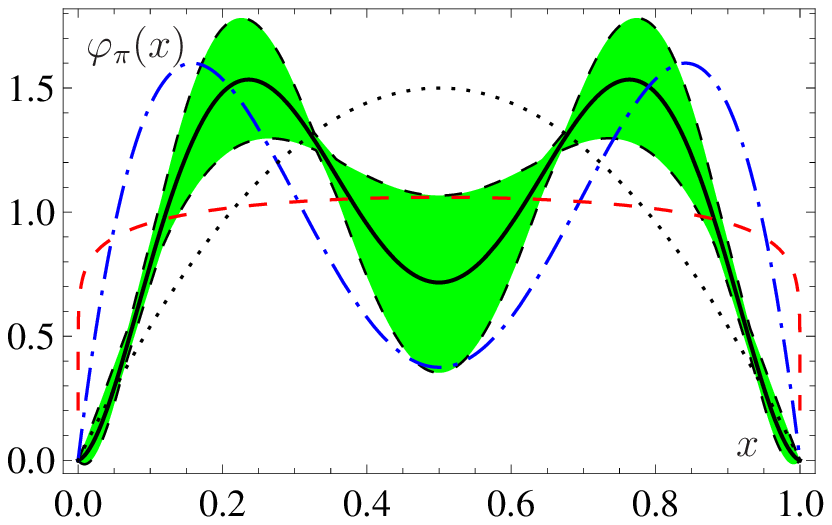}~~
            \includegraphics[width=0.45\textwidth]{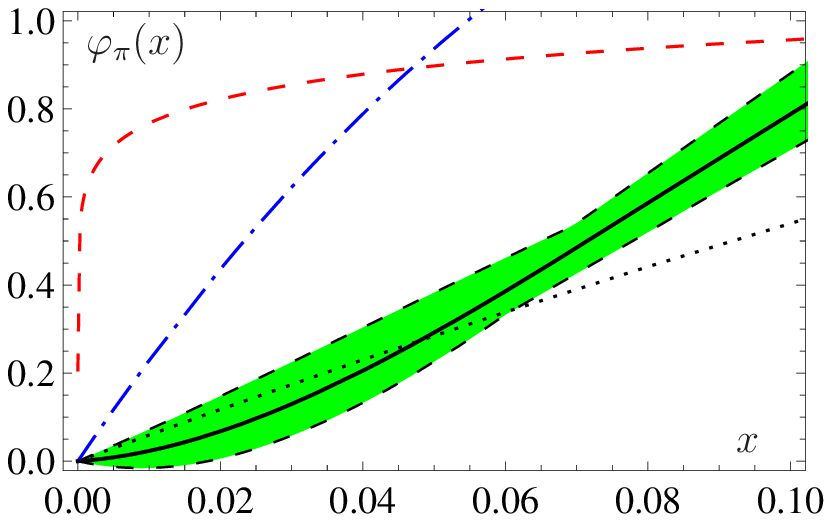}
           }
\caption{\footnotesize
Comparison of selected pion DA models.
The left panel shows their shapes, while the right panel shows a
magnified view of the endpoint region $x\sim 0$.
Solid line --- central line of the BMS bunch~\protect\cite{BMS01};
dashed-dotted (blue) line --- CZ model~\protect\cite{CZ84};
short-dashed (red) line --- flat-top DA from Eq.\ (\ref{eq:flat-DA})
with $\alpha=0.1$;
dotted line --- asymptotic DA.
All DAs are normalized at the same scale $\mu_0^2\simeq 1$~GeV$^2$.
\label{fig:pionDAmodels}}
\end{figure}

As already mentioned in the Introduction, the fine details of the pion
DA in the region around the origin $x \sim 0$ are of crucial importance.
To illustrate the differences in the end-point behavior of existing
model DAs, we show the corresponding profiles in
Fig.\ \ref{fig:pionDAmodels} (left panel).
We depict four models:
BMS~\cite{BMS01} model --- solid line;
CZ~\cite{CZ84} --- dashed-dotted (blue) line;
flat-top DA given by Eq.\ (\ref{eq:flat-DA})---short-dashed (red) line;
dotted line --- asymptotic DA.
Using the values of the derivative of the pion DA, collected in the
last column of Table~\ref{tab:all-xi-crit}, we can classify the pion
DA models according to their end-point behavior:
end-point suppressed (Asymptotic and BMS) and
end-point enhanced (CZ and flat-top (\ref{eq:flat-DA})).
A zoomed-in view of the end-point characteristics of the above DAs
is displayed in the right panel of Fig.\ \ref{fig:pionDAmodels}.

Relying upon the QCD SR approach with NLCs, developed in \cite{BMS01},
we direct our attention to the end-point behavior of the pion DA.
This will allow us to obtain predictions for the slope of the pion DA
in the end-point region with smaller errors than those obtained from
the BMS bunch~\cite{BMS01} in this region \cite{MPS10}.
Recall that the basic idea underlying the NLC parametrization of the
QCD vacuum, is that it has a domain structure, parameterized in terms
of condensates, the latter possessing a certain correlation length by
virtue of which the vacuum quarks acquire a non-zero average virtuality
$\langle k^2_q\rangle$ (see, for instance, \cite{Rad97}).
To analyze the nonlocality of the vacuum condensate, it is useful
to parameterize the lowest scalar condensate in terms of\footnote{%
In this work we use the ``fixed-point'' gauge $z^\mu A_\mu = 0$,
so that $[0,z]=1$.}
$\langle{\bar{q}(0)[0,z]q(z)}\rangle\equiv M_S(z^2)$.
This quantity can be related to the vacuum distribution function
$f_S(\alpha)$ via \cite{MR89}
\begin{eqnarray}
  M_S(z^2)
&=&
  \langle{\bar{q}q}\rangle
  \int\limits_0^\infty\!\! f_S(\alpha)\,e^{\alpha z^2/4}\,d \alpha \ ,
\label{eq:nonlocCon}
\end{eqnarray}
where $\alpha$ is the vacuum-quark virtuality.
Assuming that the vacuum quarks have a fixed virtuality $\lambda_q^2$,
one has
\begin{eqnarray}
 f_S(\alpha)
=
 \delta(\alpha-\lambda_q^2/2)\,,
  \label{eq:delta}
\end{eqnarray}
leading for the scalar-quark condensate~\cite{MR89} to the Gaussian 
model
\begin{eqnarray}
 M_S(z^2)
=
  \langle\bar{q}(0)q(0)\rangle
  {\rm e}^{-|z^{2}|\lambda_{q}^{2}/8} \ .
\label{eq:scalar-cond}
\end{eqnarray}
The parameter $\lambda_{q}^{2}$ here represents the typical quark
momentum in the QCD vacuum and it can be given the following definition
\begin{eqnarray}
  2\langle k_q^2\rangle
=
  \frac{\langle\bar{q}(0)\nabla^{2}q(0)\rangle}
       {\langle\bar{q}(0)q(0)\rangle}
\equiv
  \lambda_{q}^{2} \ .
\label{eq:lambda}
\end{eqnarray}
In our study \cite{MPS10}, presented here, we use the value
$\lambda_{q}^{2}=0.4$~GeV$^2$, which is supported by several
analyses, though values within the interval
$[0.35\div 0.45]$~GeV$^2$ are still acceptable
(see \cite{BMS02,BMS01,BM02} and references cited therein).

The QCD SRs with nonlocal condensates for the pion DA write
\cite{BMS01}
\begin{eqnarray}
  f_{\pi}^2\,\varphi_\pi(x)
  + f_{A_1}^2\,\varphi_{A_1}\!(x)\, e^{-m_{A_1}^2/M^2}
  + \int\limits_{s_0}^{\infty}
                              \rho_\text{pert}\left(x\right)
                              e^{-s/M^2}ds
=
   \int\limits_{0}^{\infty}
                           \rho_\text{pert}\left(x\right)
                           e^{-s/M^2}ds \nn\\
   + \Delta\Phi_\text{G}(x,M^2)
   + \left[\Delta\Phi_\text{S}(x,M^2)
   + \Delta\Phi_\text{V}(x,M^2)
   + \Delta\Phi_\text{T}(x,M^2)\right]_{\rm Q} \ ,
\label{eq:pionDAQCDSR}
\end{eqnarray}
where $\varphi_{A_1}$ is the $A_1$-meson DA and $f_\pi$ and
$f_{A_1}$ are, respectively, the decay constants of the pion
and the $A_1$-meson. 
Here and below we use $M^2$ notation for Borel  parameter.
Note that the $A_1$-meson state is an effective state which takes into
account both the $\pi'$ and the $a_1$ meson.
The nonperturbative input in the theoretical part of the SR (right side)
are the gluon-condensate term
$\Delta\Phi_{G}(x,M^2)$
and the quark-condensate contribution
$\left[...\right]_{\rm Q}$.
This latter contribution contains the vector-condensate term (V),
the mixed quark-gluon condensate term (T), and the scalar condensate
term (S).
The explicit expressions for the perturbative and the nonperturbative
contributions to the NLO spectral density
$\rho_\text{pert}^\text{(NLO)}\left(x\right)$
can be found in \cite{BMS01,MPS10}.
Remarkably, the first radiative correction in the spectral density in
the end-point region is of $\mathcal{O}(\alpha_s)$ and comes out too
large relative to both the zeroth-order perturbative contribution as
well as the nonperturbative parts.
For this reason, we resort in our analysis \cite{MPS10} to the
leading-order (LO) approximation
$\rho_\text{pert}^\text{(LO)}\!\left(x\right)=3x\bar x/2\pi^2$.
In fact, in order to include radiative corrections into the spectral
density (when analyzing the end-point region), one would be
obliged to resum all radiative corrections --- a formidable task for
the future.

Among the nonperturbative terms, the scalar-quark condensate provides
the largest smooth contribution at the origin $x\sim 0$, notably,
\begin{eqnarray}
  \Delta\Phi_\text{S}(x,M^2)
&=&
  18 A_S (\Phi' x + \Phi'' x^2 +O(x^3))
\label{eq:4Qterm.x}
\end{eqnarray}
with the coefficients
$
 \Ds A_S
=
 \frac{8\pi\alpha_S}{81}\langle\bar qq\rangle^2\,,
$
\begin{eqnarray}
  \Phi'
\simeq
   \int_0^\infty\!\frac{f_S(\alpha)}{\alpha^2}\,d\alpha
=
   \langle{\bar{q}q}\rangle^{-1}\!\!\int_0^\infty\!\!
   z^2 M_S(z^2)\,dz^2
~~\text{and}~~
  \Phi''
\approx
  \Phi'M^2\int_0^{1/2}\!\!\!~f_S(M^2 t)\frac{t}{1-t}\,dt
\ .
\label{eq:4Qterm}
\end{eqnarray}
Inspection of the expression for the first coefficient $\Phi'$
reveals that the end-point behavior of the pion DA is directly
related to the behavior of the scalar-quark condensate at
large/moderate distances.

\begin{figure}[h!]
\centerline{\includegraphics[width=0.5\textwidth]{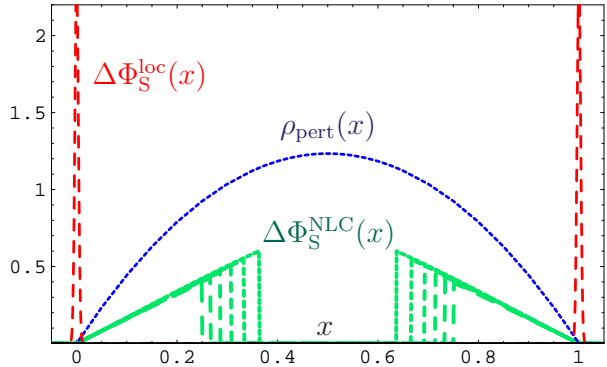}}
\caption{\footnotesize
 Comparison of the leading nonperturbative contribution in the local
 ($\Delta\Phi_\text{S}^\text{loc}(x)$)
 and the nonlocal
 ($\Delta\Phi_\text{S}^\text{NLC}(x)$)
 condensate models along with the perturbative contribution
 ($\rho_\text{pert}(x)$).
\label{fig:4Qloc.nonloc.pert}}
\end{figure}

On the other hand, all nonperturbative terms in the \emph{local}
condensate model ($\lambda_q^2=0$) are concentrated exactly at the
endpoints ($\Delta\Phi_\text{S}^\text{loc}(x)\sim\delta(x)$), as shown
in Fig.\  \ref{fig:4Qloc.nonloc.pert} in terms of the
scalar-condensate contribution.
In contrast, the nonlocal condensate leads to an end-point-suppressed
nonperturbative contribution.

In using the SR (\ref{eq:pionDAQCDSR}) in order to study the end-point
behavior of the pion DA, we have to find an appropriate characteristic
that is capable of describing the slope of the pion DA at the origin.
Because the nonperturbative terms are strongly concentrated at the
endpoints (see Fig.~\ref{fig:4Qloc.nonloc.pert}), viz.,
$\Delta\Phi_\text{V}\sim x\,\delta'(\Delta-x)$,
$\Delta\Phi_\text{G}\sim \delta(\Delta-x), \ldots$
with $\Delta=\lambda_q^2/(2 M^2)$,
the best way to take into account all these contributions is to
investigate the integral characteristics of the pion DA.
For this reason we invented in \cite{MPS10} the ``integral
derivative'', which will occupy us in the next section.

If we apply the Gaussian model (\ref{eq:scalar-cond}), then only the
four-quark condensate $\Delta\Phi_\text{S}\sim x\,\theta(\Delta-x)$
contributes in the end-point region --- see
Fig.\ \ref{fig:4Qloc.nonloc.pert}--- without leading to singularities.
But, if we assume a behavior of the various condensates differing from
the delta-ansatz model, and use, for instance, for the scalar-quark
condensate a smooth model like (\ref{eq:smoothmodel}) (which implies
a decay at large distances not slower than the exponential decay),
then the other nonperturbative terms (V), (G), and (T) contribute
only small amounts in the small-$x$ region.
Therefore, these terms can be neglected if one is only interested in
deriving the simplest characteristic, i.e., the derivative of the pion
DA at the origin $\varphi_\pi'(0)$, considered in
Sec.\ \ref{subsec:dif-SR}.

\section{``Integral'' sum rules}
\label{subsec:int-SR}

The key features of the applied SRs are collected in Table
\ref{tab:QCD-SR-works}.
Because the QCD SRs were developed with the aim to study the integral
characteristics of the pion DA, most approaches appeal to the moments
\begin{equation}
    \langle \xi^{N} \rangle_{\pi}
        \equiv
    \int_{0}^{1} (2x-1)^{N} \varphi_{\pi}(x)\,dx\, ,
\label{eq:moments}
\end{equation}
where $\xi\equiv 2x-1$.
Once these moments are known, they can be used in order to reverse
engineer the pion DA, with a precision depending upon the influence
of the magnitude of the discarded higher-order moments.
The zero-order moment leads to a SR for the decay constant, studied
long ago by Shifman, Vainshtein and Zakharov~\cite{SVZ}.
The second-order term of the Gegenbauer expansion was determined by
Chernyak and Zhitnitsky~\cite{CZ84} from the standard SRs with local
condensates ($\lambda_q^2=0$).
Still higher-order coefficients were computed \cite{BMS01} using
QCD SRs with NLCs.
It was shown there that one can de facto resort to the first
two Gegenbauer coefficients $a_2$ and $a_4$ because the values of
$a_i$ with $i=6, 8, 10$ were calculated and found to be negligible.
On that basis, the authors of \cite{BMS01} obtained a bunch of
two-parameter dependent pion DAs, among them also the BMS model,
mentioned earlier, and an independent SR for the inverse 
moment~(\ref{eq:inv-mom}).
The inverse moment itself is a rather good, though not sufficient
(see for arguments~\cite{BMS05lat}), indicator for the end-point
behavior of the pion DA, as we can see from the fifth column of
Table \ref{tab:all-xi-crit}.

\begin{table}[thb]\vspace*{-3mm}
\caption{\footnotesize
Comparison of results for the moments of the pion DA obtained within
different approaches which use QCD SRs with local or nonlocal
condensates.
\label{tab:QCD-SR-works}
}\vspace*{+3mm}
\centerline{
\begin{tabular}{|c||c|c|c||c|} \hline \pha
Approach         & Characteristics                     & Accuracy  & Condensate & Result                  \\ \hline\pha
SVZ~\cite{SVZ}   & $\langle \xi^{0}\rangle$            &  LO       & local      & $f_\pi$                 \\ \hline\pha
CZ~\cite{CZ84}   & $\langle \xi^{2N}\rangle$, $N=0,~1$ &  LO       & local      & $f_\pi,\,a_2$           \\ \hline\pha
BMS~\cite{BMS01} & $\langle x^{-1}\rangle$, $\langle \xi^{2N}\rangle$,
                   $N=0,1,\ldots,5$                    & NLO       & nonlocal   & $f_\pi,\,a_2,\,a_4$,
                                                                                  $\langle x^{-1}\rangle$ \\ \hline\pha
Here~\cite{MPS10}& $\varphi_\pi'(0)$,
                   $[D^{(\nu)}\varphi_\pi](x)$
                                                       &  LO       & nonlocal  & $\varphi_\pi'(0)$,
                                                                                 $[D^{(\nu)}\varphi_\pi](x)$\\ \hline
\end{tabular}
}
\end{table}

In our work \cite{MPS10} we probed the endpoint region of the pion DA
in another way which employs an averaged  ``integral'' derivative.
As we explained in Sec.\ \ref{sec:pi-slope}, the usefulness of these
derivatives follows from the fact that they can be applied to QCD
SRs which may even contain singular terms.
With the help of the averaging procedure, one can take into account
all these contributions.
To expose the usefulness of this procedure, let us construct the
following sequence of average derivatives obeying the condition
$\varphi(0)=0 $: standard derivative, finite-difference derivative, and 
generalized inverse moment given by
\begin{eqnarray}
  [D^{(0)}\varphi](x)
=
  \varphi'(x)\,,~~~
  [D^{(1)}\varphi](x)
=
  \varphi(x)/x\,,~\text{and}~~
  [D^{(2)}\varphi](x)
=
  \frac{1}{x}\int\limits_0^x\frac{\varphi(y)}{y}\,dy 
\label{eq:int-dev012}
\end{eqnarray}
respectively.
The integral derivatives $D^{(\nu)}$, generalizing this sequence, 
were defined in \cite{MPS10} by means of the following expression:
\begin{eqnarray}
  [D^{(\nu)}\varphi](x)
=
  \frac{1}{x} \int\limits_0^x\!\!\varphi(y) f(y,\nu-2,x)\,dy \ ,
  f(y,\nu,x)
=
  \frac{\theta(x-y)}{\Gamma(\nu+1)\, y}
  \left(\ln \frac{x}{y}\right)^\nu
  ~\text{at}~ \varphi(0)=0 \ .
  \label{eq:D.nu.x}
\end{eqnarray}
When $\nu=2$ and $x=1$, the integral derivative coincides with the
inverse moment of the pion DA:
$[D^{(2)}\varphi](1)=\langle x^{-1} \rangle_{\pi}$.
Assuming that the Taylor expansion of $\varphi(x)$ at $x=0$ exists,
one finds from~(\ref{eq:D.nu.x})
\begin{eqnarray}
  [D^{(\nu)}\varphi](x)
  =
  \varphi'(0)+\varphi''(0)\frac{x}{2!2^{\nu-1}}
  + O\left(\frac{x^2}{3^{\nu-1}}\varphi^{(3)}(0)\right) \ ,
\label{eq:D.k.x.series}
\end{eqnarray}
which is valid for any  \emph{real} $\nu$, as we explained in detail
in~\cite{MPS10}.
From the above equation, one can see that the defined operator
$D^{(\nu)}$ reproduces at small $x$ and/or large $\nu$
the derivative of $\varphi(x)$ at the origin $x=0$.

\begin{figure}[h!]
\centerline{\includegraphics[width=0.5\textwidth]{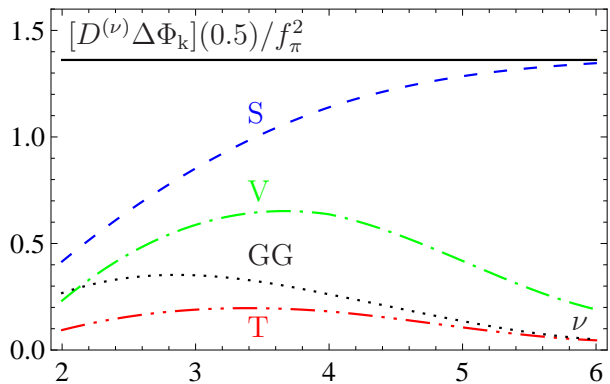}
         }
\caption{\footnotesize
Mutual comparison of the nonperturbative contributions
(k=S,~V,~T, and GG) to the SR (\ref{eq:pionDAQCDSR:Dk}) for the integral
derivatives $[D^{(\nu)}\varphi_\pi](x)$ of the pion DA and with the value
$18 A_S \Phi'/f_\pi^2=72 A_Sf_\pi^{-2}\lambda_q^{-4}$ (horizontal solid line)
of the scalar-quark contribution to the SR (\ref{eq:DpionDAQCDSR})
for the standard derivative $\varphi_\pi'(0)$.
The scalar-quark condensate term (k=S) is shown as a dashed (blue) line.
The other terms are the vector-quark condensate term (k=V)  ---
dashed-dotted (green) line, the  mixed quark-gluon condensate term (k=T)
--- dashed-dotted-dotted (red) line, and the gluon-condensate term (k=GG)
--- dotted (black) line.
\label{fig:nonpert.terms}}
\end{figure}

By applying the operator $[D^{(\nu+2)}]$ on both sides of the QCD
SR given by (\ref{eq:pionDAQCDSR}), we obtain a new SR for
$[D^{(\nu+2)}\varphi_\pi](x)$, viz.,
\begin{equation}
\begin{split}
&   f_{\pi}^2\,[D^{(\nu)}\varphi_\pi](x)
  + f_{A_1}^2\,{\rm e}^{-m_{A_1}^2/M^2}[D^{(\nu)}\varphi_{A_1}](x)
  + \int\limits_{s_0}^{\infty}
      [D^{(\nu)}\rho_\text{pert}]\left(x\right) {\rm e}^{-s/M^2}ds
  \\
& =
  \int\limits_{0}^{\infty}
  [D^{(\nu)}\rho_\text{pert}]\left(x\right) {\rm e}^{-s/M^2}ds
  + [D^{(\nu)}\Delta\Phi_\text{G}](x,M^2)
  + [D^{(\nu)}\Delta\Phi_\text{V}](x,M^2)  \\
& ~~~~~~~~~~~~~~~~~~~~~~~~~~~~~~~~~~~~~~~
  + [D^{(\nu)}\Delta\Phi_\text{T}](x,M^2)
  + [D^{(\nu)}\Delta\Phi_\text{S}](x,M^2)
  \, .
\end{split}
\label{eq:pionDAQCDSR:Dk}
\end{equation}
The detailed analysis of this SR can be found in~\cite{MPS10}.
Here we only discuss the nonperturbative terms depicted in
Fig.~\ref{fig:nonpert.terms}.
The dominant contribution to the integral derivative
$[D^{(\nu)}\varphi_\pi](0.5)$ of the pion DA stems from the
scalar-quark condensate (S), while the vector-condensate (V),
the mixed quark-gluon condensate (T), and the gluon-condensate (GG)
contributions are comparatively less important.
It turns out that the image of the operator $D^{(\nu)}$ for 
$\nu \geq 6$ is numerically very close to the result obtained with the 
differentiation method (see next section) --- for any $x$.

\begin{figure}[b!]
\centerline{\includegraphics[width=0.51\textwidth]{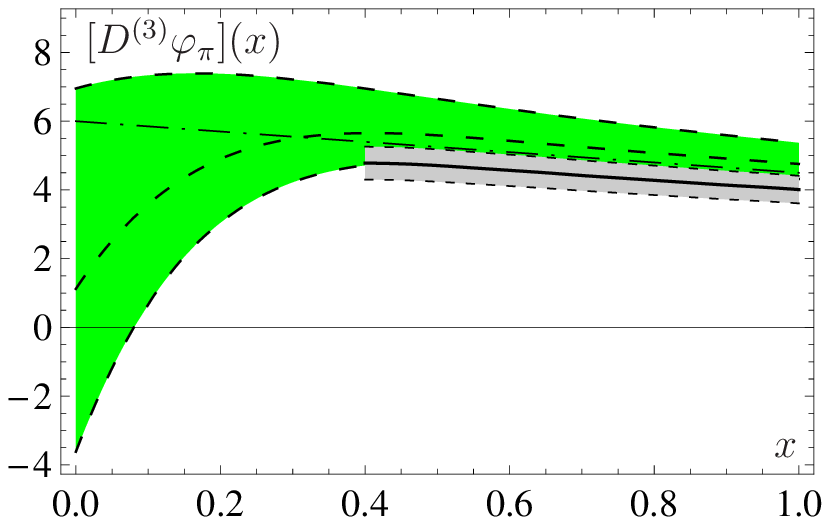}
           ~\includegraphics[width=0.48\textwidth]{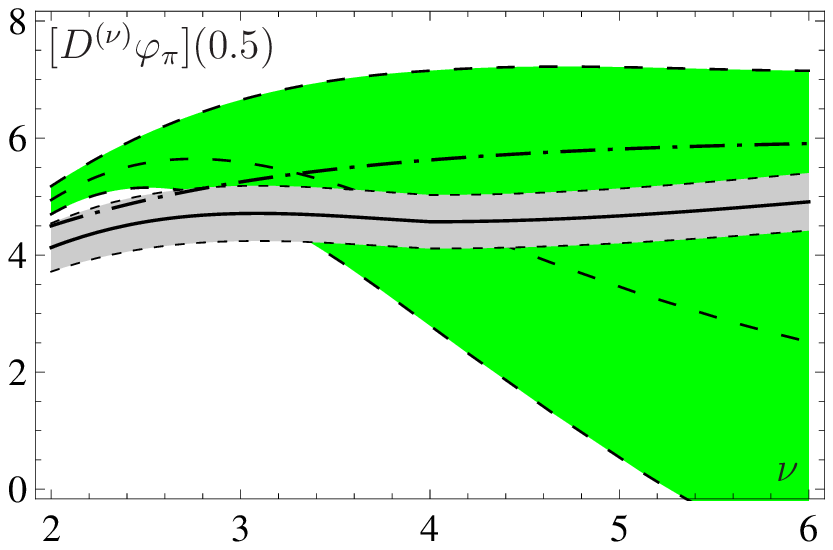}   
         }
\caption{\footnotesize
  $x$-dependence (left panel) and $\nu$-dependence (right panel) of
  $[D^{(\nu)}\varphi_\pi](x)$ shown for the BMS bunch of pion DAs
  \protect\cite{BMS01} (shaded green band within long-dashed lines)
  in comparison with the SR result (\ref{eq:pionDAQCDSR:Dk})
  (narrow gray strip) in both panels.
  The left panel shows the predictions for $[D^{(3)}\varphi_\pi](x)$,
  whereas those for
  $[D^{(\nu)}\varphi_\pi](0.5)$ are presented in the right panel.
  The dashed-dotted line denotes the asymptotic result
  $[D^{(\nu)}\varphi^\text{Asy}](x)=6-3x/2^{\nu-2}$.
\label{fig:DDDA012}}
\end{figure}

For large $\nu$, the (S)-term dominates and is close to the value
obtained by the standard derivative illustrated in Fig.\
\ref{fig:nonpert.terms} by the horizontal line, while all other
condensate terms disappear.
Thus, the integral SR (\ref{eq:pionDAQCDSR:Dk}) becomes close to
the differential one to be considered in the next section.
For this reason, we analyze the constructed SR
(\ref{eq:pionDAQCDSR:Dk}) for $\nu\in[2,6]$ and $x>0.4$ and present
the results in Fig.\ \ref{fig:DDDA012} in terms of a solid line that
is inside the light gray strip bounded by the short-dashed lines.
For the sake of comparison, the predictions for the asymptotic
DA (dashed-dotted line) and the BMS DA bunch --- obtained in the
NLC SR analysis of Ref.\ \cite{BMS01} --- (shaded band limited by
long-dashed lines) are also shown.
From this figure we see that our SR estimates for
$[D^{(\nu)}\varphi^\text{SR}_\pi](x)$ agree fairly well with the BMS
model --- see also Table \ref{tab:all-derivatives}.
This table shows estimates for the third-order integral derivative of
the pion DA for $x=0.5$, using (i) the sum rule given by Eq.\
(\ref{eq:pionDAQCDSR:Dk}) and (ii) the pion DA models we discussed
above; in addition, results pertaining to flat-type DAs are also
included.

First, we compare the QCD SR result, obtained from
(\ref{eq:pionDAQCDSR:Dk}), with what one finds with the flat-type DA
models.
Consider first the flat-top DA model defined by
\begin{eqnarray}
  \varphi_\text{flat}(x)
&=&
   \frac{\Gamma(2(\alpha+1))}{\Gamma^2(\alpha+1)}x^\alpha(1-x)^\alpha
   \ .
\label{eq:flat-DA}
\end{eqnarray}
This model was invented in \cite{Rad09} and attempts to describe
the BaBar data via a logarithmic behavior with $Q^2$.
For the value $\alpha=0.1$, one finds
$[D^{(3)}\varphi^\text{flat}](0.5)=227$, which is much
larger and far outside the range of values extracted from our
SR (\ref{eq:pionDAQCDSR:Dk}).

As a second option, we consider a particular flat-type pion DA which
is provided by the AdS/QCD correspondence in the holographic
approach --- see, for instance, Refs.\ \cite{BT07,KL08,AGN08}.
In that case one has $\alpha=0.5$ yielding
$[D^{(3)}\varphi^\text{hol}](0.5)=14$.

On the other hand, the CZ model --- being also endpoint-enhanced ---
yields third-order integral derivatives which are incompatible with
the values derived from our SR (\ref{eq:pionDAQCDSR:Dk}) (cf.\
Table~\ref{tab:all-derivatives}).
Note that a similar statement also applies to the pion DA
proposed in \cite{WH10}, which employs a Brodsky-Huang-Lepage
ansatz for the $\mathbf{k}_\perp$-dependence of the pion wave
function --- see Table~\ref{tab:all-derivatives}.
The main message from this table is that the SR for the integral
derivative of the pion DA is fulfilled by the BMS bunch,
whereas flat-type DAs and the CZ model DA have no overlap with
the estimated range of values.
By contrast, the pion DA model proposed in \cite{WH10} --- though it
provides a similarly large integral derivative like the CZ DA --- has a
usual derivative at the origin which is zero due to the strong
exponential suppression of this DA in the small vicinity of the origin.
Finally, the model DA defined by Eq.\ (\ref{eq:flat-DA}) has no
derivative at the origin while the integral derivative is well-defined.
As we see from these examples, the integral derivative allows one to
compare a broader range of the pion DA values in the end-point region
than the standard one.

\begin{table}[thb]\vspace*{-3mm}
\caption{\footnotesize
Results for the third-order integral derivative at the value $x=0.5$
and such for the standard derivative of the pion DA, using different
SR approaches (first three rows) and pion DA models (six last rows).
\label{tab:all-derivatives}}\vspace*{+3mm}
\begin{tabular}{|c|l|c|c|} \hline
\pha  ~~~& Approach/Model                           & ``Integral derivative'' $[D^{(3)}\varphi_\pi](0.5)$
                                                                        & Derivative $\varphi'_\pi(0)$ \\ \hline
\pha   1 & Integral SR (\ref{eq:pionDAQCDSR:Dk})    & $4.7 \pm 0.5$     & $5.5 \pm 1.5$                \\ \hline
\pha   2 & Differential SR (\ref{eq:DpionDAQCDSR})  & ---               & $5.3 \pm 0.5$                \\ \hline
\pha   3 & SR (\ref{eq:DpionDAQCDSR}) with
           smooth NLC (\ref{eq:smoothmodel})        & ---               & $7.0 \pm 0.7$                \\ \hline\hline
\pha   4 & BMS bunch~\cite{BMS01}                   & $5.7\pm 1.0$      & $1.7 \pm 5.3$                \\ \hline
\pha   5 & Asymptotic DA                            & $5.25$            & $6$                          \\ \hline\hline
\pha   6 & CZ DA~\cite{CZ82}                        & $15.1$            & $26.2$                       \\ \hline
\pha   7 &    DA from \cite{WH10}                   & $14$              & $0$                          \\ \hline
\pha   8 & AdS/QCD DA \cite{BT07}                   & $14$              & $\gg 6$                      \\ \hline
\pha   9 & flat-top DA
           (Eq.\ (\ref{eq:flat-DA}), $\alpha=0.1$)  & $227$             & $\gg 6$                      \\ \hline
\end{tabular}
\end{table}
It is worth mentioning that the usual derivative
$\varphi_\pi'(0)$ of the pion DA encapsulates the key
characteristics of the pion DA at small $x$.
This quantity can be extracted from
$[D^{(\nu)}\varphi^\text{SR}_\pi](x)$,
which can be derived from the SR (\ref{eq:pionDAQCDSR:Dk}) employing
different values of $\nu$ and $x$.
The results are shown in Fig.~\ref{fig:DDDA012}.
To determine $\varphi_\pi'(0)$ one can first use two terms of the
Taylor expansion (\ref{eq:D.k.x.series}) and then subtract the second
derivative for which the asymptotic value
$\varphi''_\pi(0)= -12(6)$ is taken.
That yields the estimate
$\varphi'_\pi(0)=5.5 \pm 1.5$ for any $0.4<x$ and $2 \leq \nu \leq 6$.

\section{Differential sum rules}
\label{subsec:dif-SR}

Another way to study the behavior of the pion DA in the small-$x$
region is provided by the differentiation of the SR
(\ref{eq:pionDAQCDSR}), which yields
\begin{eqnarray}
  f_{\pi}^2\,\varphi_\pi'(0,M^2)
=
  \frac{3}{2\pi^2}M^2\left(1-{\rm e}^{-s_0/M^2}\right)
  + 18 A_S \Phi'
  -f_{A_1}^2\,\varphi_{A_1}'\!(0)\,{\rm e}^{-m_{A_1}^2/M^2} \ .
\label{eq:DpionDAQCDSR}
\end{eqnarray}
We shall evaluate this SR for the threshold value $s_0=2.61$~GeV$^2$,
recalling that we are employing a LO expression for the spectral
density.
As it was shown in our recent work~\cite{MPS10}, only the
four-quark condensate survives and gives a contribution to the SR
defined by Eq.~(\ref{eq:4Qterm}).
>From this equation we see that the nonperturbative contribution
to the SR is mainly due to the scalar-quark condensate
at large/moderate distances $z^2\gtrsim 4/\langle k_q^2\rangle$.
Employing the delta-ansatz
$
 f_S(\alpha)
=
 \delta(\alpha-\lambda_q^2/2)
$
the nonperturbative contribution to SR (\ref{eq:DpionDAQCDSR}) reduces
to the following simple expression
\begin{eqnarray}
  \Phi'
=
  \Phi_\text{delta}'
&=&
  \frac{4}{\lambda_q^4} \ .
\label{eq:delta-dev}
\end{eqnarray}

Having fixed the ingredients of the SR, we are now able to consider
the implications of using the smooth model for the quark-virtuality
distribution in the differential SR.
Despite the usefulness of the Gaussian model, we have to take into
account the possibility that the scalar-quark condensate may behave
differently at asymptotically large distances.
Indeed, there are indications from heavy-quark effective theory
\cite{Rad91} that it could decay exponentially.
Note that in order to ensure the existence of the vacuum matrix element
$\langle{\bar{q}(D^2)^Nq}\rangle$, the quark-virtuality
distribution $f_S(\alpha)$ should decrease faster than any power
$1/\alpha^{N+1}$ as $\alpha\to\infty$ \cite{MR89}.
Following this reasoning, a two-tier model for $f_S$ was proposed in
\cite{BM02,BM96} which has a smooth dependence on the quark virtuality
$\alpha$.
Hence, one has
\begin{eqnarray}
  f_{S}(\alpha;\Lambda,n,\sigma)
&=&
  \frac{\left(\sigma/\Lambda\right)^{n}}{2K_n(2\Lambda\sigma)}\,
  \alpha^{n-1} e^{-\Lambda^2/\alpha-\alpha\,\sigma^2} \ ,
\label{eq:smoothmodel}
\end{eqnarray}
where $K_n(z)$ is the modified Bessel function.
This so-called ``smooth model'', depends on two parameters
$\Lambda$ and $\sigma$
which serve to take into account the long- and the short-distance
behavior of the nonlocal condensates.
For large distances $|\,z|=\sqrt{-z^2}$, this model leads to the
asymptotic form of the scalar quark NLC
\begin{eqnarray}
  M_S(z^2)
&\stackrel{|\,z|\to\infty}{\longrightarrow}&
  \langle{\bar{q}q}\rangle|\,z|^{-(2n+1)/2}e^{-\Lambda|z|}
  \frac{2^{(2n-1)/2}\sqrt{\pi}\,\sigma^n}{\sqrt{\Lambda}\,
  K_n(2\Lambda\sigma)} \ .
\label{eq:smooth-x}
\end{eqnarray}

Let us now consider briefly this model --- referring for technical
details to \cite{BM02,BM96} --- and set $n=1$, whereas the second
parameter $\Lambda=0.45$ GeV can be taken from the QCD SRs for the
heavy-light meson transition in heavy quark effective theory
\cite{Rad91,Rad94}.
The two parameters $n$ and $\Lambda$ are responsible for the large-$z$
behavior of the scalar-quark condensate, cf.\ Eq.\ (\ref{eq:smooth-x}).
The third parameter $\sigma^2=10$ GeV$^{-2}$ is defined in terms
of the parameters $n, \Lambda$, and $\lambda_q^2$ via the following
equation
\begin{eqnarray}
  \int_0^\infty\!\!\!\alpha\,
  f_{S}(\alpha;\Lambda,n,\sigma)\,d\alpha
=
  \frac{\Lambda}{\sigma}
  \frac{K_{n+1}(2\Lambda\sigma)}{K_n(2\Lambda\sigma)}
=
  \frac{\lambda_q^2}{2} \ ,
\label{eq:sigma-FS}
\end{eqnarray}
which we are going to evaluate for the value of the nonlocality
parameter $\lambda_q^2=0.4$~GeV$^2$.
The main effect of using a smooth model for the quark-virtuality
distribution in comparison to the Gaussian form,
${f_S(\alpha)=\delta(\alpha-\lambda_q^2/2)}$,
is the increase of the nonperturbative contribution to the SR,
induced this way, entailing the relation
\begin{eqnarray}
  \Phi_\text{smooth}'
&=&
  \int_0^\infty\!
  \frac{f_{S}(\alpha;\Lambda,n,\sigma)}{\alpha^2}\,d\alpha
=
  \frac{\sigma^2}{\Lambda^2}
  \frac{K_{n-2}(2\Lambda\sigma)}{K_{n}(2\Lambda\sigma)}
>
  \Phi_\text{delta}' \ .
\label{eq:smooth-1}
\end{eqnarray}

We studied the SR (\ref{eq:DpionDAQCDSR}) for this model adopting the
following values of its parameters:
$
 f_{S}(\alpha; \Lambda=0.45~\text{GeV}, n=1,
\sigma^2=10~\text{GeV}^{-2})
$.
It turns out that the average value of the derivative
$\varphi_\pi'(0,M^2)$ in the fiducial Borel interval is
$\varphi_\pi'(0)=7.0(7)$,
meaning that the nonperturbative contribution
$\Phi_\text{smooth}'$,
obtained from the smooth model, is approximately two times larger
than the analogous contribution $\Phi_\text{delta}'$ for the
delta ansatz
$\Phi_\text{smooth}'\approx 2.3\,\Phi_\text{delta}'$.
Appealing to the relation (\ref{eq:4Qterm}), it seems reasonable to
conclude that choosing a model for the condensate that has a slower
decay at large distances (small $n$ or $\Lambda$), may induce an
increase of the nonperturbative contribution to the SR
(\ref{eq:pionDAQCDSR}) and, hence, entail an increase of the value
$\varphi_\pi'(0)$ as well.
The option of having a condensate model with a faster
decay at large distances (large $n$ or $\Lambda$), leads to a
decrease of the nonperturbative contribution to the SR
(\ref{eq:pionDAQCDSR}) and therefore to a decrease of the value
$\varphi_\pi'(0)$.
To facilitate the comparison of these two distinct possibilities
for the scalar-quark condensate, we give in the last column of
Table \ref{tab:all-derivatives} the values of the (usual) pion DA
derivative at $x\simeq 0$, using various SR approaches (first
three rows) and selected pion DA models (last six rows).

\section{Conclusions}
\label{sec:concl}

In the present work we proposed a direct way to access the end-point
characteristics of the pion DA in the QCD SR approach with nonlocal
condensates.
To characterize the slope of the pion DA at the origin, we introduced
\cite{MPS10} a suitable operator for integral derivatives that allows
us to describe a broader range of the pion DA in the end-point region
than the standard derivative.
Moreover this operator provides the possibility to include all
condensate terms of the QCD SRs with nonlocal condensates.

Our results have been presented in the second and the third
column of Table \ref{tab:all-derivatives} for the range of values of
the integral and the standard derivatives of the pion DA, respectively.
The same table contains also the values of the derivatives of some
characteristic pion DA models, viz., the BMS, the CZ, the asymptotic,
and the flat-top DA given by Eq.\ (\ref{eq:flat-DA}) with $\alpha=0.1$.
The dependence of the integral derivative on its parameters $x$ and
$\nu$ is displayed in Fig.\ \ref{fig:DDDA012}.

From Table \ref{tab:all-derivatives}, we conclude that the
differential (\ref{eq:DpionDAQCDSR}) and the integral
(\ref{eq:pionDAQCDSR:Dk}) SRs agree rather well with each other.
The integral and the standard derivatives of the pion DA,
based on these new SRs, give smaller values than the asymptotic one and
overlapping with the range of values determined with the BMS bunch
of pion DAs \cite{BMS01}, while there is no agreement with the CZ
DA and the considered flat-top model.

It is worth remarking that, employing the integral and the differential
QCD SR (\ref{eq:pionDAQCDSR})--- which employ the minimal Gaussian model 
for the nonlocal condensates --- cannot be satisfied by flat-type pion 
distribution amplitudes.
Using a physically motivated exponential decay model by means of
expression (\ref{eq:smoothmodel}), leads to a higher value of the slope
of the pion DA at the origin $\varphi_\pi'(0)$, though it is still
much smaller than the corresponding value of the flat-type pion
distribution amplitudes.

Some final comments:
It turns out that the nonperturbative content in the differential SR
for $\varphi_\pi'(0)$ is mainly due to the scalar-quark condensate,
a feature valid also for the slope obtained with the integral SR.
To be specific, the scalar-quark condensate term is proportional to
the second inverse moment (\ref{eq:4Qterm}) of the distribution
$f_S(\alpha)$ of the vacuum-quark virtuality and is determined by the
behavior of the quark condensate at large/moderate distances
of the vacuum quarks.
By virtue of (\ref{eq:4Qterm}), we may conclude that, adopting a
model for the scalar-quark condensate that has a slower decay at
large distances, entails an increase of the nonperturbative
contribution to the SRs (\ref{eq:pionDAQCDSR:Dk} and
\ref{eq:DpionDAQCDSR}), so that also the value of the pion DA slope
defined via the integral and the standard derivatives increases.

The presented analysis shows that it is difficult to reconcile
QCD sum rules, and related techniques and features of the QCD
nonperturbative vacuum in terms of nonlocal condensates, with
flat-type pion DAs.

\section{Acknowledgments}

We would like to thank Alexander Bakulev for stimulating discussions
and useful remarks.
A.V.P. is indebted to Prof. Maxim Polyakov for the warm hospitality
extended to him at Bochum University.
He also wishes to thank the Ministry of Education and Science of the
Russian Federation (``Development of Scientific Potential in Higher
Schools'' projects: No.\ 2.2.1.1/1483 and No.\ 2.1.1/1539), the
Russian ``Dynasty'' Foundation  for a research scholarship, and
the DAAD Foundation (Germany) for a research grant.
This work received partially support from the Heisenberg--Landau
Program under Grants 2009 and 2010, the Russian Foundation for
Fundamental Research (Grants No.\ 07-02-91557, No.\ 08-01-00686, and
No.\ 09-02-01149), and the BRFBR-JINR Cooperation Program, contract
No. F06D-002.


\end{document}